\documentclass[aps,pra,twocolumn,groupedaddress,showpacs,floatfix]{revtex4}

\usepackage{color,soul}

\usepackage{epsfig}
\usepackage{amssymb}
\usepackage{amsmath}

\usepackage{rotfloat}
\usepackage{theorem}
\usepackage{epic}
\usepackage{graphics}
\usepackage{rotating}
\usepackage{placeins}
\usepackage{lscape}
\usepackage{longtable}
\usepackage{dcolumn}
\usepackage[usenames,dvipsnames]{pstricks}

\usepackage{epsfig}
\usepackage{hyperref}

\begin{document}

\newcolumntype{.}{D{.}{.}{-1}}


\newcommand{\del}{\partial}
\newcommand{\beq}{\begin{equation*}}
\newcommand{\eeq}{\end{equation*}}
\newcommand{\be}{\begin{equation}}
\newcommand{\ee}{\end{equation}}
\newcommand{\beqa}{\begin{eqnarray}}
\newcommand{\eeqa}{\end{eqnarray}}
\newcommand{\bea}{\begin{eqnarray}}
\newcommand{\eea}{\end{eqnarray}}
\newcommand{\req}[1]{Eq.\,(\ref{#1})}

\newcommand{\bra}{\langle}
\newcommand{\ket}{\rangle}
\newcommand{\Tr}{{\rm Tr}\,}
\newcommand{\tr}{{\rm Tr}\,}
\newcommand{\s}{\sigma}
\newcommand{\w}{\omega}
\newcommand{\reci}[1]{\frac{1}{#1}}
\newcommand{\half}{\frac{1}{2}}
\newcommand{\emdash}{\hspace{1pt}---\hspace{1pt}}
\newcommand{\volint}[1]{\int \frac{d^4{#1}}{(2\pi)^4} \;}
\newcommand{\volthree}[1]{\int_0^{#1_F} \frac{d^3{#1}}{(2\pi)^3} \;}
\newcommand{\fixminus}{\raisebox{1.5pt}{\mathunderscore}\hspace{0.5pt}}

\newcommand{\putat}[3]{\begin{picture}(0,0)(0,0)\put(#1,#2){#3}\end{picture}}

\hypersetup{
    colorlinks=true,       
    linkcolor=red,          
    citecolor=green,        
    filecolor=magenta,      
    urlcolor=blue           
}

\urlstyle{same}


\title{Non-Perturbative Relativistic Calculation of the Muonic Hydrogen Spectrum}



\author{J.~D.~Carroll} 
\email{jcarroll@physics.adelaide.edu.au}
\author{A.~W.~Thomas}

\affiliation{Centre for the Subatomic Structure of Matter (CSSM),
School of Chemistry and Physics, University of Adelaide, SA 5005, Australia}

\author{J.~Rafelski}
\affiliation{Departments of Physics, University of
  Arizona, Tucson, Arizona, 85721 USA}

\author{G.~A.~Miller}
\affiliation{University of Washington, Seattle, WA 98195-1560 USA}


\date{\today}

\begin{abstract}
We investigate the muonic hydrogen $2P_{3/2}^{F=2}$ to
$2S_{1/2}^{F=1}$ transition through a precise, non-perturbative
numerical solution of the Dirac equation including the finite-size
Coulomb force and finite size vacuum polarization. The results are
compared with earlier perturbative calculations of (primarily) Borie,
Martynenko, and
Pachucki~\cite{Borie:2004fv,Borie:1982ax,Borie:1975xb,Martynenko:2004bt,Martynenko:2006gz,Pachucki:1996zza};
and experimental results recently presented by Pohl {\it et
  al.}~\cite{Pohl:2010zz}, in which this very comparison is
interpreted as requiring a modification of the proton charge radius
from that obtained in electron scattering and electronic hydrogen
analyses. We find no significant discrepancy between the perturbative
and non-perturbative calculations, and present our results as
confirmation of the perturbative methods.
\end{abstract}

\maketitle

\section{Introduction}\label{sec:intro}

\begin{figure}
\centering \includegraphics[width=0.5\textwidth]{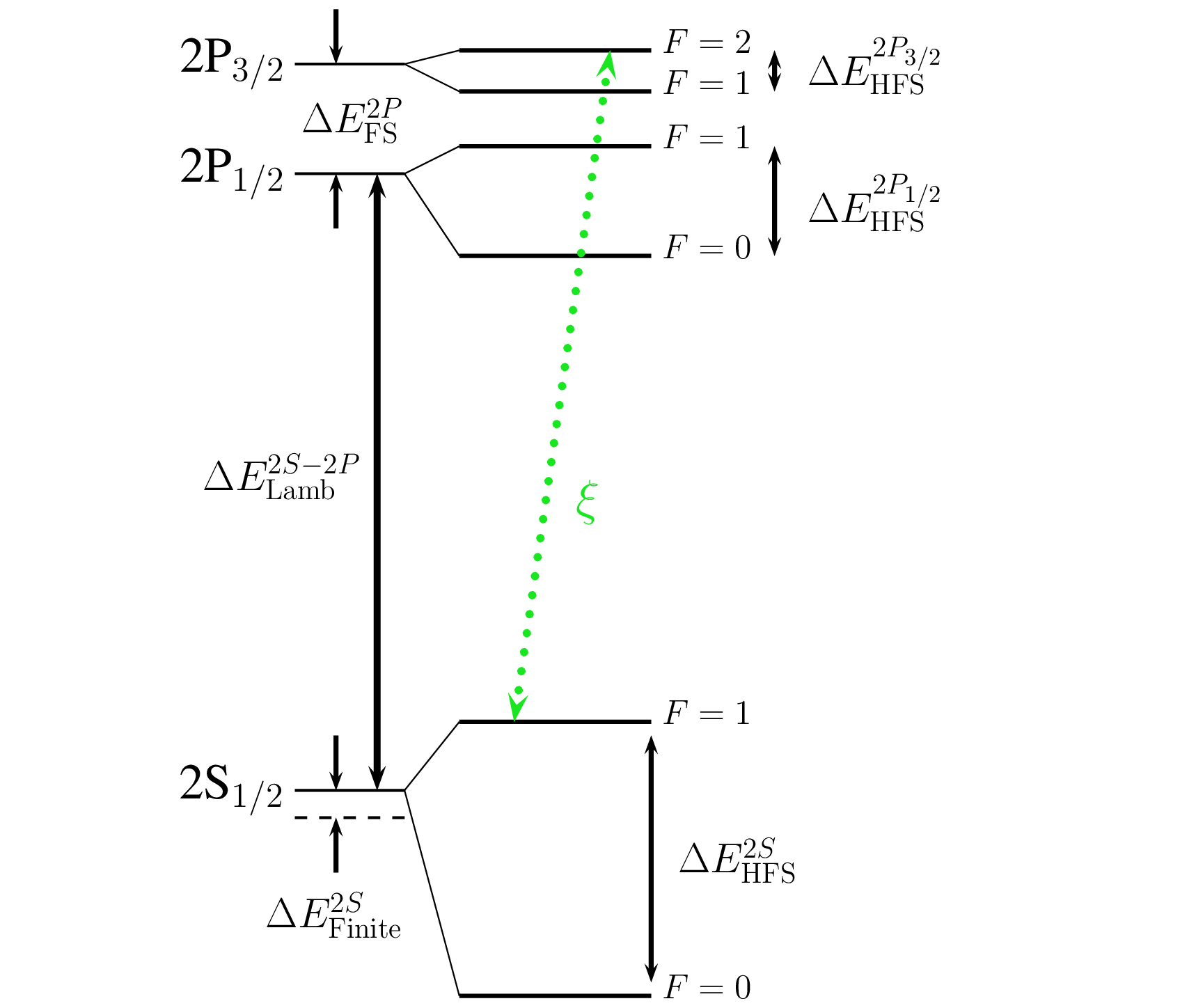}
\caption{(Color online) Muonic hydrogen L-shell spectrum, showing the
  finite proton-size correction, Lamb shift splitting, fine structure,
  and hyperfine structure. The $2S_{1/2}^{F=1}$ to $2P_{3/2}^{F=2}$
  transition measured by Pohl {\it et al.}~\cite{Pohl:2010zz} is shown
  as the green dotted line (marked $\xi$).\protect\label{fig:spectrum}}
\end{figure}

The precision measurement of the Lamb shift transition energy between
the $2P_{3/2}^{F=2}$ and $2S_{1/2}^{F=1}$ states of muonic hydrogen by
Pohl {\it et al.\/}~\cite{Pohl:2010zz}, see Fig.~\ref{fig:spectrum},
has created considerable interest because of a 0.31~meV discrepancy
with the value predicted by theoretical calculations (speficically
those discussed in Ref.~\cite{Pohl:2010zz};
Borie~\cite{Borie:2004fv,Borie:1982ax,Borie:1975xb} and
Martynenko~\cite{Martynenko:2004bt,Martynenko:2006gz} along with many
others~\cite{Eides:2000xc,Rafelski:1977vq,Friar:1978wv,Fricke:1969fh}). This
Lamb shift splitting of $\cal{O}$(206)~meV is dominated by the lowest
order QED vacuum polarization, and obtains a significant contribution
from the finite size of the proton. On selecting and combining the
perturbative predictions for the corresponding contributions to the
measured transition, Pohl {\it et al.} produce a cubic equation
relating their experimentally measured energy shift to the theoretical
prediction, and arrive at
\bea
\nonumber
206.2949(32)~{\rm meV} &=& 206.0573(45) \qquad\qquad\qquad\\
\nonumber
&& \ - 5.2262 \bra r_p^2 \ket^{1/2} \\
&& \ \ \ + 0.0347 \bra r_p^2
\ket^{3/2}~{\rm meV},
\eea

the only physically-meaningful solution of which implies a proton rms
charge radius of $r_p \equiv \sqrt{\bra r_p^2\ket} = 0.84184(67)~{\rm
  fm}$ which differs from the consensus CODATA~\cite{Mohr:2008fa}
value of $r_p = 0.8768(69)~{\rm fm}$ by 4.9 standard deviations.

Such a large modification of a basic electromagnetic property of the
proton suggests that either there may be an as yet unrecognised
problem in several other experimental efforts (such as the electronic
hydrogen spectroscopy and scattering experiments which primarily lead
to the CODATA value~\cite{Mohr:2008fa}) or in the QED
calculations~\cite{Miller:2011yw,Hill:2011wy,Cloet:2010qa}, or
alternatively that some new physics (beyond the Standard Model)
contributes to this transition
energy~\cite{TuckerSmith:2010ra,Jaeckel:2010yy,Batell:2011qq}. With
respect to the QED calculations, we note that the predominant
theoretical approach involves perturbation theory applied to the
solutions of the non-relativistic Schr\"odinger
equation~\cite{Borie:2004fv,Martynenko:2006gz,Martynenko:2004bt}. Since
the effects of finite size and the vacuum polarization potential are
quite large at short distance, it seems important to verify by
explicit calculation that the perturbative treatment is indeed
adequate at the level quoted.

We therefore calculate the transition energy relevant to
the aforementioned experiment using the relativistic Dirac equation to
describe the muon wave function non-perturbatively. We take care to
control the numerical errors in the calculations and to quantify any
differences from the perturbative non-relativistic approach. In our
work we extend considerably the earlier work by
Borie~\cite{Borie:2004fv,Borie:1982ax}.

In the sections following, we discuss the nature of the transition and
contributing physical effects, as well as the method by which we
calculate the energies corresponding to the various eigenstates. We
summarize any discrepancies with respect to previous work.
%
%
%

%
\section{Muonic Hydrogen Spectrum}\label{sec:measurement}

After muon capture by hydrogen, about 1\% of the muons reach the
metastable $2S$-state. This state is activated by laser excitation from
$2S_{1/2}^{F=1}$ to $2P_{3/2}^{F=2}$ and the signature that the
laser energy is well tuned is the appearance of a prompt E1 transition
X-Ray of a muon from the $2P_{3/2}^{F=2}$ state of muonic hydrogen to
the 1S$_{1/2}^{F=1}$ state. Given the resonance method used to
establish the energy of the Lamb shift, there appears to be no doubt
about the remarkable experimental result of Pohl {\it et
  al.}~\cite{Pohl:2010zz}.

The L-shell level scheme explored in this experiment is depicted in
Fig.~\ref{fig:spectrum} for the point-Coulomb potential of magnitude
\be \label{eq:pointC}
V_C(r) = -\frac{Z\alpha}{r}.
\ee
The $2S_{1/2}$ and $2P_{1/2}$ states are degenerate even in the
Coulomb-Dirac theory. Since the muon orbit is 200 times smaller than
the corresponding electron orbit, the S-states probe the point charge
smearing effect of vacuum polarization, which is the dominant
component in the Lamb shift, $\Delta E_{\rm Lamb}^{2S-2P}$, in
Fig.~\ref{fig:spectrum}. Of similar nature is a smaller but
significant effect for hydrogen ($Z=1$) arising from the finite proton
charge radius, denoted $\Delta E_{\rm Finite}^{2S}$ in
Fig.~\ref{fig:spectrum}. While the vacuum polarization effect
increases the binding of the S$_{1/2}$ states pulling these `down' in
Fig.~\ref{fig:spectrum}, the finite charge distribution acts in the
opposite direction.

The spin-orbit fine structure splitting of the $2P_{1/2}$ and
$2P_{3/2}$ eigenstates is predicted by the Dirac equation. It is
denoted by $\Delta E_{\rm FS}^{2P}$ in Fig.~\ref{fig:spectrum}. As
the measured transition does not involve the $2P_{1/2}$ state, we
shall only calculate the relevant energy levels for the purpose of
comparison with perturbative results.

The spin-spin coupling of the muon to the proton adds hyperfine
splitting to the spectrum. For the case of the $2S_{1/2}$ eigenstate,
this leads to hyperfine eigenstates with total angular momenta $F=0$
and $F=1$. The muon involved in the measured transition decays to the
$2S_{1/2}^{F=1}$ state, and thus we address an accurate determination
of the energy of this state and the corresponding splitting $\Delta
E_{\rm HFS}^{2S}$. Similarly, the hyperfine states are labeled by
$F=0$ and $F=1$ for the $2P_{1/2}$ case, and $F=1$ and $F=2$ for the
$2P_{3/2}$ case. The hyperfine splitting energies are denoted here by
$\Delta E_{HFS}^{{2P}_{1/2}}$ and $\Delta E_{HFS}^{{2P}_{3/2}}$ for
the corresponding states. The muon in the measured transition decays
from the $2P_{3/2}^{F=2}$ eigenstate, so we require an accurate
determination of the energy of this state. We note that the
muon-proton tensor force does lead to some mixing of the two F=1
states but we will not make a new calculation of this effect.

%
\section{Numerical Method}\label{sec:numerical method}

To calculate the theoretical energy difference corresponding to the
measured transition, previous authors have primarily used perturbation
theory with non-relativistic wave functions, including in the
effective interaction terms describing the various relativistic
corrections. In order to calculate the perturbative effect on the
energy produced by an operator, $\delta V$, additional to the Coulomb
potential,
\be 
V_{\rm eff} = -\frac{Z\alpha}{r} + \delta V , 
\ee 
we require a wave function to integrate over. If we follow the
methods of previous authors, we can use exact Schr\"odinger
wave functions for states with quantum numbers $n, \ell, m$ and the
lowest order correction to the energy is
\be \label{eq:perturb}
\Delta E_{\rm V'}^{n\ell m} \approx 
\int_0^\infty \phi_{\rm Schr\ddot{o}d.}^{n\ell m\ \dagger}(r)\delta V\, 
\phi_{\rm Schr\ddot{o}d.}^{n\ell m}(r)  d^3r.
\ee

An alternate approach, which we choose here, is to use the Dirac
equation with the appropriate potential in order to calculate the
perturbed wave functions. This approach is known to be a specific
limit approximation to the two-particle Bethe-Saltpeter
equation~\cite{Eides:2000xc} .

We consider the muon wave function in state $\alpha$ to be a spinor
$\psi_\alpha(\vec{r\, })$
\be
\psi_\alpha(\vec{r\, }) = \begin{pmatrix}
    g_\alpha(r) \chi_{\kappa}^\mu(\hat{r}) \\[2mm]
    -if_\alpha(r) \chi_{-\kappa}^\mu(\hat{r})
  \end{pmatrix}  = \begin{pmatrix}
    {\displaystyle \frac{G_\alpha(r)}{r}}\ \chi_{\kappa}^\mu(\hat{r}) \\[2mm]
    {\displaystyle \frac{-iF_\alpha(r)}{r}}\ \chi_{-\kappa}^\mu(\hat{r})
  \end{pmatrix},
\ee
normalised to unity, such that the probability is
\be
\int |\psi_\alpha|^2\; d^3r = \int_0^\infty r^2 \left[g_\alpha(r)^2+f_\alpha(r)^2
\right]\; dr = 1,
\ee
noting that $\chi_{\kappa}^\mu$ are eigenfunctions of the total
angular momentum operator (consisting of a combination of spherical
harmonics and Pauli spinors) satisfying
\be
\int \chi_\kappa^{m\dagger}\chi_{\kappa'}^{m'}\;
d\hat{r} = \delta_{\kappa\kappa'}\delta_{mm'}.
\ee

The separation of center of mass motion is not exact for a
relativistic two body system. To lowest order in the ratio of muon to
proton mass we use the reduced mass $\mu$ in place of the muon mass in
the Dirac Equation
\be
\mu = \frac{M_p m_\mu}{M_p+m_\mu} \, .
\ee
Along with further recoil corrections treated in perturbation
theory, this should provide a very accurate description of the system.

The binding of the muon in this system is extremely weak (on the scale
of rest energy) and as such the eigenvalue, $\epsilon_\alpha$, for
each state calculated using the Dirac equation is approximately equal
to the reduced mass $\mu$. In order to precisely calculate the
variance from this value, we rewrite the Dirac equation to incorporate
the shift of eigenvalue down by the reduced mass, such that the
eigenvalue we are now solving for is $\lambda_\alpha = \epsilon_\alpha
- \mu$, corresponding to the binding energy.

The Dirac equation we wish to solve is therefore given by 
%
\be
\label{eq:dirac}
\displaystyle{\frac{d}{dr}}\begin{pmatrix}G_\alpha(r)\\[2mm] F_\alpha(r)\end{pmatrix} 
= \left(\begin{array}{cc}
-{\displaystyle \!\!\frac{\kappa_\alpha}{r}} &\!\!\!\! \lambda_\alpha\! + 2\mu - V \\[2mm]
-\lambda_\alpha\! + V & \!\!\!\!{\displaystyle \frac{\kappa_\alpha}{r}}
\end{array}\right)
\begin{pmatrix}G_\alpha(r)\\[2mm] F_\alpha(r)\end{pmatrix},
\ee
%
where the value of $\kappa_\alpha$ is specific to each eigenstate, namely
\begin{eqnarray*}
{1S}_{1/2}: \kappa =& -1,\quad {2S}_{1/2}: \kappa =& -1, \\
{2P}_{1/2}: \kappa =& +1,\quad {2P}_{3/2}: \kappa =& -2.
\end{eqnarray*}

In order to integrate \req{eq:dirac}, we supply an initial guess for
the eigenvalue $\lambda_\alpha$ and appropriate boundary conditions
for the upper and lower components of the wave function at small and
large radii. We then integrate from each limit towards a central
matching point. The normalised discontinuity in the wave function
integrated from each limit is used as a measure of the inaccuracy of
the eigenvalue and a refined estimate is calculated. This process is
iterated until $\lambda_\alpha$ changes by less than the required
tolerance.

%
\section{Quality Control}\label{sec:qc}

To convince ourselves that our method is self-consistently accurate,
we check the accuracy of our procedure using several methods. The
unperturbed point-Coulomb Dirac eigenvalues are known analytically to
be
\be 
\label{eq:exactDirac}
\lambda_\alpha = \epsilon_\alpha - \mu =
\mu\left[1+\frac{Z^2\alpha^2}{(n_\alpha-|\kappa_\alpha|+
\sqrt{\kappa_\alpha^2-Z^2\alpha^2})^2}\right]^{-\half}\!\!\!\!\!-\mu, 
\ee
where $n_\alpha$ denotes the principle quantum number for state
$\alpha$. We first ensured that were are able to reproduce these
values within a reasonable compute-time. For the $2S_{1/2}$ wave
function, we reproduce the analytic result to within 10~neV, the
$2P_{1/2}$ eigenstate to within 40~neV and the $1S_{1/2}$ and
$2P_{3/2}$ eigenstates to within 10~peV, using quad-precision Fortran,
a sufficiently large grid size, and sufficiently small grid spacing.

This test does not assure that solutions for a realistic case\emdash
including the finite size of the proton as well as finite size vacuum
polarization\emdash converges with same accuracy. For this reason we employ
the virial theorem test for our solutions (refer to
Ref.~\cite{Rafelski:1977vq} for further details) by calculating the
reduced eigenvalue as
\be \label{eq:virial}
\lambda_\alpha = \bra \psi_\alpha | \mu(\beta-1) + V(\vec{r\, }) +
\vec{r\, }\cdot \vec{\nabla}V(\vec{\, r}) | \psi_\alpha \ket.
\ee
 
The virial theorem provides a far more stringent test of the accuracy
of the muon wave function near the origin, where $|\vec{\nabla}V|$ is
greatest. We find that the eigenvalues calculated using
\req{eq:exactDirac} and \req{eq:virial} for the $2S_{1/2}$ wave
function differ by 180~neV for a point-Coulomb potential, and 450~neV
the finite-Coulomb plus finite-vacuum polarization potentials
discussed in Sec.~\ref{sec:finite}. We therefore conservatively take
our errors to be less than $\pm 500$~neV. Propagating this error, we
find that in principle we could determine the required proton rms
charge-radius to within approximately 0.05~am ($5\times
10^{-5}$~fm). This should be sufficient to provide a reliable,
independent test of the accuracy of the perturbative approach, however
we note that the determination of the proton rms charge radius cannot
be performed to this precision as the error in that analysis is
dominated by experimental error in the determination of the transition
energy.

%
\section{$2S_{1/2}$--$2P_{1/2}$ Lamb Shift}\label{sec:lamb}

The Lamb shift is the splitting of the otherwise degenerate
$2S_{1/2}$ and $2P_{1/2}$ eigenstates attributed to the vacuum
polarization potential $V_{\rm VP}$, which for a point source is given
in \cite{Pachucki:1996zza} as 
\be
\label{eq:VP}
V_{\rm VP}(r) =
-\frac{Z\alpha}{r}\frac{\alpha}{3\pi}\int_4^{\infty}\frac{e^{-m_{e}qr}}{q^2}\ 
\sqrt{1-\frac{4}{q^2}}\left(1+\frac{2}{q^2}\right)dq^2,
\ee
where $m_e$ is the electron mass. We can calculate the effect that
this has on the eigenvalues by assuming that this potential is a small
perturbation of the Coulomb potential, and thus using
Eq.~(\ref{eq:perturb}) we find
\be \label{eq:borielamb}
\Delta E_{\rm Lamb}^{nlm} \approx \int_0^\infty V_{\rm VP}(r)\, 
|\Psi_{\rm Schr\ddot{o}d.}^{nlm}(r)|^2 d^3r\, ,
\ee
to which we must also add higher order perturbation theory, 
relativistic, recoil, and radiative
corrections and generally higher-order (in $\alpha$) corrections.

Alternatively\emdash and more accurately\emdash we can calculate the
shift in eigenvalues using converged Dirac wave functions in response
to the combined effect of the Coulomb and vacuum polarization potentials. 
In this case we
simply take the difference between the converged eigenvalues for the
$2S_{1/2}$ and $2P_{1/2}$ eigenstates calculated in the presence of
point-Coulomb and point-vacuum polarization potentials
\be
\Delta E_{\rm Lamb}^{2S-2P} = \lambda_{2P_{1/2}} -
\lambda_{2S_{1/2}}\, = 205.1706(5)~{\rm meV},
\ee

Care must be taken when comparing this calculation to that of
\req{eq:borielamb} since our calculation includes relativistic
corrections, which are treated as corrections to \req{eq:borielamb} in
Ref.~\cite{Borie:2004fv}. A summary of this comparison and the
calculated values for the Lamb shift are given in
Table~\ref{tab:lamb}, where we note that the perturbative and
non-perturbative calculations are found to be in good agreement. For
this table and those that follow, we refer to various iterations of
our calculations in which we compute the wave function in the presence
of point-Coulomb (C); finite (size nucleus)-Coulomb (FC); point vacuum
polarization (VP); and finite (size nucleus) vacuum polarization (FVP)
potentials. The dependence on $\bra r_p^2\ket^n$ is extracted in each
case by fitting the energy shifts calculated at various values of $r_p
\equiv \bra r_p^2\ket^{1/2}$ to a cubic of the form given in
Eq.~(\ref{eq:cubic}) for the case of a Coulomb-only potential, and
with the addition of a term proportional to 1 when including the
vacuum polarization (to account for the $2S$--$2P$ splitting). The values
listed in the column `Pohl {\it et al.}' of Table~\ref{tab:lamb} have
their origins in references 1--5, 11, 12, 14--17, and 19--25 of
Ref~\cite{Pohl:2010zz} and are taken to be reliable values.

%
\begin{center}
\begin{table*}[bt]
\footnotesize
\caption{Contributions to the $2S$--$2P$ Lamb shift with comparison to
  values presented in Pohl {\it et al.}~\cite{Pohl:2010zz} which
  themselves are selected values from various theoretical
  sources\emdash references 1--5, 11, 12, 14--17, and 19--25 of Pohl
  {\it et al.}. Values are all in meV. Errors in the Dirac
  calculations are taken to be $\pm 500$~neV as per
  Section~\ref{sec:qc}. We refer to various iterations of our
  calculations in which we compute the wave function in the presence
  of point-Coulomb (C); finite-Coulomb (FC); point vacuum polarization
  (VP); and finite vacuum polarization (FVP) potentials. The
  dependence on $\bra r_p^2\ket^n$ is extracted in each case by
  fitting the energy shifts calculated at various values of $r_p$ to a
  cubic of the form given in Eq.~(\ref{eq:cubic}) for the case of a
  Coulomb-only potential, and with the addition of a term proportional
  to 1 when including the vacuum polarization. The listed corrections
  are already included in our Dirac calculations, namely lines 3 and 5
  of Table~1 in Ref.~\cite{Pohl:2010zz} and the nuclear size
  contributions of Table~2 from that reference. All further
  corrections to both the perturbative calculation and our calculation
  are contained in `Remaining Corrections' which in this case
  encompasses all remaining contributions of Table~1 and radiative
  correction of Table~2 of
  Ref.~\cite{Pohl:2010zz}. \label{tab:lamb}\vspace{2mm}}
\begin{tabular}{lccc}
&& \\[-3mm]
\hline
\hline\\[-2mm]
Contribution & Pohl {\it et al.} & \phantom{abcde} & Present Work \\[1mm]
\hline\\[-2mm]

Dirac ($V = V_{\rm C} + V_{\rm VP}$) & & & 205.1706 \\[1mm]
Dirac ($V = V_{\rm FC}$) & & & \phantom{200.000} -5.2000 $\bra r_p^2\ket$ + 0.0350 $\bra r_p^2\ket^{3/2}$ \\[1mm]
Dirac ($V = V_{\rm FC} + V_{\rm VP}$) & & & 205.1706 - 5.2169 $\bra r_p^2\ket$ + 0.0353 $\bra r_p^2\ket^{3/2}$ \\[1mm]
\hline
&& \\[-2mm]
Dirac ($V = V_{\rm FC} + V_{\rm FVP}$)\phantom{abcde} & & & 205.1822 - 5.2519 $\bra r_p^2\ket$ + 0.0546 $\bra r_p^2\ket^{3/2}$ \\[1mm]

Relativistic one loop VP & 
205.0282 &   \\[1mm]

Polarization insertion in & \\
two Coulomb lines & 
0.1509   &   \\[1mm]


Finite size effects &
-5.1987 $\bra r_p^2\ket$ + 0.0347 $\bra r_p^2\ket^{3/2}$ \\[1mm]

\hline
&& \\[-2mm]
Subtotal: & 205.1791 - 5.1987 $\bra r_p^2\ket$ + 0.0347 $\bra r_p^2\ket^{3/2}$ & & 205.1822 - 5.2519 $\bra r_p^2\ket$ + 0.0546 $\bra r_p^2\ket^{3/2}$\\[1mm]

\hline
&& \\[-2mm]

Remaining Corrections & \multicolumn{3}{c}{0.8782 - 0.0275~$\bra r_p^2\ket$} \\[1mm]

\hline
&& \\[-2mm]
Total: & 206.0573 - 5.2262 $\bra r_p^2\ket$ + 0.0347 $\bra r_p^2\ket^{3/2}$
& & 206.0604 - 5.2794 $\bra r_p^2\ket$ + 0.0546 $\bra r_p^2\ket^{3/2}$ \\[1mm]

\hline
\hline

\end{tabular}
\end{table*}
\end{center}
%

%
\section{Proton Finite-Size Corrections}\label{sec:finite}

The perturbative leading-order contributions associated with the
finite size of the proton arise from consideration of the proton-form
factor. These are introduced by Borie~\cite{Borie:1975xb} and
considered in Friar~\cite{Friar:1978wv}, Section VI below
Eq.~(64b). These terms appear in Pohl {\it et al.}~\cite{Pohl:2010zz}
as quoted from Ref.~\cite{Borie:2004fv}, and are given by
\be
\Delta E_{\rm Finite} =
-\frac{2Z\alpha}{3}\left(\frac{Z\alpha\mu}{2}\right)^3\left[\bra
  r_p^2\ket - \frac{Z\alpha\mu}{2}\bra r_p^2\ket^{3/2}\right]. \\
\ee
 
To calculate this effect in our fully relativistic, non-perturbative
calculation, we consider the replacement of the point-Coulomb
potential with the finite-size Coulomb potential in
Eq.~(\ref{eq:dirac})
\be \label{eq:finiteC}
V_{C}(r) = -\frac{Z\alpha}{r} \to
-Z\alpha\int\frac{\rho(r')}{|\vec{r}-\vec{r\, }'|}\, d^3r',
\ee
where $\rho(r)$ is the proton charge-distribution (or more accurately,
the Fourier transform of the Sachs electric form-factor). In response
to concerns that the shape of the form factor may significantly
influence the theoretical calculations~\cite{DeRujula:2010zk}, we have
studied the dependence of the finite-size correction on the form of
this term (always normalised to unity) and this will be summarized in
an upcoming publication (Ref.~\cite{Carroll:2011de}), though the
dependence on the choice of charge-distribution\emdash whether it be
exponential, Yukawa, or Gaussian in form\emdash appears to be
extremely weak.\par

The exponential form for the charge-distribution,
normalised to unity such that ${\displaystyle \int \rho(r)\, d^3r =
  1}$ is given by
\be \label{eq:CD}
\rho(r) = \frac{\eta}{8\pi} e^{-\eta r}; \quad \eta =
\sqrt{12/\bra r_p^2\ket}\, .
\ee

We calculate the Lamb shift by taking the difference between the
appropriate eigenvalues calculated using the Dirac equation with the
potential given by Eq.~(\ref{eq:finiteC}) with the charge-distribution
given by Eq.~(\ref{eq:CD}) for various values of $\bra r_p^2\ket$. We
then interpolate the energy shifts and fit the data to a cubic of the
form
\be \label{eq:cubic}
f(x) = A \bra r_p^2\ket + B \bra r_p^2\ket^{3/2},
\ee 
which provides the relevant parameterization. The $\bra r_p^2\ket^n$
dependence in the presence of an exponential finite-sized Coulomb
potential and point vacuum polarization potential
\be
V(r) = -Z\alpha\int\frac{\rho(r')}{|\vec{r}-\vec{r\, }'|}\, d^3r' + V_{\rm VP}(r),
\ee
is given by 
\be
\Delta E_{\rm Finite} = 205.1706 - 5.2169 \bra r_p^2\ket + 0.0353 \bra
r_p^2\ket^{3/2}~{\rm meV}.
\ee

A further important effect of the finite size of the proton arises
through the convolution of the vacuum polarization potential
(Eq.~(\ref{eq:VP})) with the proton charge-distribution. This leads to
the replacement of the point vacuum polarization potential by
\be
V_{\rm VP}(r) \to
-\frac{2Z\alpha^2}{3\pi}\int\frac{\rho(r')}{|\vec{r}-\vec{r\, }'|}Z_0(|\vec{r}-\vec{r\, }|)\, d\tau',
\ee
where we use the expression given in Ref.~\cite{Fricke:1969fh}
\be
Z_n(|\vec{r\, }|) = \int_1^\infty
e^{-\frac{2}{\lambdabar}|\vec{r}|\xi} \left(1+\frac{1}{2\xi^2}\right)\frac{(\xi^2-1)^{\half}}{\xi^n\xi^2}d\xi,
\ee
and where $\lambdabar$ denotes the electron Compton wavelength
(divided by $2\pi$), $\lambdabar = 386.15926459~{\rm fm}$. When
discussing this potential, it should be assumed that we are using a
normalised exponential charge-distribution.
We once again calculate the eigenvalues using
various values of $\bra r_p^2\ket$ in the charge-distribution and fit
the resulting energies to a cubic (as per Eq.~(\ref{eq:cubic})),
except that in this case the vacuum polarization induces the Lamb
shift, and we must include a term proportional to 1. Thus we find
\be
\Delta E_{\rm Finite} = 205.182 - 5.2519 \bra r_p^2\ket + 0.0546 \bra
r_p^2\ket^{3/2}~{\rm meV}, 
\ee

which is the expression which is compared to the perturbative
calculation in Table~\ref{tab:lamb}. We note that the finite-vacuum
polarization induces a small but non-trivial shift, and that the
results are otherwise essentially the same as those of Pohl {\it et
  al.}~\cite{Pohl:2010zz}.

%
\section{$2P$ Fine Structure}\label{sec:2pFS}

The ${\cal O}(Z\alpha)^4$ perturbative $2P$ fine structure splitting is
calculated in Ref.~\cite{Martynenko:2006gz} to be
\be \Delta E_{FS}^{2P} =
\frac{\mu^3(Z\alpha)^4}{32m_\mu^2}\left(1+\frac{2m_\mu}{m_p}\right),
\ee
along with higher-order corrections. Taking this
splitting as the difference between the converged eigenvalues of the
$2P_{1/2}$ and $2P_{3/2}$ eigenstates gives 
\be
\Delta E_{FS}^{2P} = \lambda_{{2P}_{3/2}} - \lambda_{{2P}_{1/2}}\, ,
\ee
which we can also calculate in the presence of the various
potentials. For the case of an exponential finite-Coulomb potential
with finite vacuum polarization, the $2P$ fine structure splitting is 
\be
\Delta E_{FS}^{2P} = 8.4206(5)~{\rm meV}.
\ee
 
A comparison of this value with perturbative calculations of
Borie~\cite{Borie:2004fv} is presented in Table~\ref{tab:2Pfs}. The
effect of the finite-size Coulomb potential (as compared to the point
case) is negligible at the level of errors of our
calculation. Similarly, the effect of the finite-size vacuum
polarization is also negligible at our level of errors. The vacuum
polarization itself increases the fine structure splitting by
$5~\mu$eV. We note that the perturbative and non-perturbative
calculations are in perfect agreement to the level of errors presented
here.

%
\begin{center}
\begin{table}[bt]
\footnotesize
\caption{Contributions to the $2P$ fine-structure splitting with
  comparison to values found in Borie~\cite{Borie:2004fv}. Subscripts
  are defined in Table~\ref{tab:lamb}. Values are all in meV. Errors
  in the Dirac calculations are taken to be $\pm 500$~neV as per
  Section~\ref{sec:qc}. The listed correction
  (Uehling/vacuum polarization) is already included in our Dirac
  calculations. All further corrections to both the perturbative
  calculation and our calculation are contained in `Remaining
  Corrections' which are detailed in Table~II of
  Ref.~\cite{Borie:2004fv}. Finite-size effects in either the Coulomb
  or vacuum polarization potentials provide no shift above the level
  of errors here, as expected for P-states. The perturbative
  calculation prediction is reproduced within
  errors.\label{tab:2Pfs}\vspace{2mm}}
\begin{tabular}{l..}
&& \\[-3mm]
\hline
\hline\\[-2mm]
Contribution & \multicolumn{1}{c}{Borie} & \multicolumn{1}{c}{Present Work} \\[1mm]
\hline\\[-2mm]

Dirac ($V = V_{\rm C}$) & 8.4156 & 8.4156 \\[1mm]
Dirac ($V = V_{\rm FC}$) & & 8.4156 \\[1mm]
Dirac ($V = V_{\rm C} + V_{\rm VP}$) & & 8.4206 \\[1mm]
Dirac ($V = V_{\rm FC} + V_{\rm VP}$) & & 8.4206 \\[1mm]
\hline
&& \\[-2mm]
Dirac ($V = V_{\rm FC} + V_{\rm FVP}$) & & 8.4206 \\[1mm]

Uehling (VP) &
0.0050 & \\[1mm]

\hline
&& \\[-2mm]
Subtotal &
8.4206 & 8.4206 \\[1mm]

\hline
&& \\[-2mm]
Remaining Corrections & \multicolumn{2}{c}{-0.06852} \\[1mm]

\hline
&& \\[-2mm]
Total: & 8.3521 & 8.3521 \\[1mm]

\hline
\hline

\end{tabular}
\end{table}
\end{center}
%

%
\section{Hyperfine Structure}\label{sec:HFS}
 
The hyperfine structure is a measure of the $\vec{\ell}\cdot\vec{\s}$
coupling. Following the lead of Ref.~\cite{weissbluth1978atoms}, the
appropriate Hamiltonian is given by
\be
\label{eq:hamiltonian}
{\cal H} =
2\beta_\mu\gamma\hbar\frac{\ell(\ell+1)}{j(j+1)}\bigg\bra\frac{1}{r^3}\bigg\ket
  \mathbf{I}\cdot\mathbf{J} +
  \frac{16\pi}{3}\beta\gamma\hbar|\psi(0)|^2\, \mathbf{I}\cdot\mathbf{S},
\ee
comprising a dipole term and a contact term, for which the following
definitions apply for the muon Bohr magneton $\beta_\mu$; proton Bohr
magneton $\beta_p$; and proton gyromagnetic ratio $\gamma$
\be
\beta_\mu = \sqrt{\alpha}/2m_\mu,\quad \beta_p = \sqrt{\alpha}/2M_p,\quad \gamma = 2(1+\kappa)\beta_p \, .
\ee
Here $\kappa = 1.792847351$ is the proton anomalous magnetic
moment. $\psi(0)$ represents the muon wave function at the centre
of the proton. We now investigate the two terms of
Eq.~(\ref{eq:hamiltonian}) separately.

%
\subsection{$2S_{1/2}$ Hyperfine Structure}\label{sec:2sHFS}

There exist several methods by which the $2S$ hyperfine structure can be
calculated. The perturbative $2S$ hyperfine structure calculated in
Ref.~\cite{Martynenko:2004bt} is given by
\be
\Delta E_{HFS}^{{2S}_{1/2}} = \frac{1}{3}(Z\alpha)^4\frac{\mu^3}{m_\mu m_p}(1+\kappa).
\ee
For $\ell=0$ the contact term in the Hamiltonian
(Eq.~(\ref{eq:hamiltonian})) is non-zero, while the dipole term
vanishes;
\be
E_{HFS}^{2S} = \frac{16\pi}{3}\beta_\mu\gamma\hbar|\psi(0)|^2 \bra Fm_F |
\mathbf{I}\cdot\mathbf{S} | Fm_F \ket\, ,
\ee
where $|Fm_F\ket$ is the eigenfunction belonging to $\mathbf{F} =
\mathbf{I} + \mathbf{J}$, such that
\be
 \bra Fm_F |
\mathbf{I}\cdot\mathbf{S} | Fm_F \ket = \half\left[F(F+1)-\frac{3}{2}\right] .
\ee
Thus, the splitting between the $2S$ $F=0$ and $F=1$ hyperfine levels is
given by
\be \label{eq:2shfs_weisskopf}
\Delta E_{HFS}^{2S(F=1-F=0)} =
\frac{16\pi}{3}\beta_\mu\gamma\hbar|\psi(0)|^2,
\ee

We note an important, relevant typographical correction; In
Ref.~\cite{weissbluth1978atoms} Eq.~(18.2-17b), the sign should be
positive and the second 9 in the denominator should not appear.

The value of the $2S$ hyperfine splitting, as calculated using
Eq.~(\ref{eq:2shfs_weisskopf}) with the wave function calculated with
the Dirac equation in the presence of the combined exponential
finite-Coulomb and finite vacuum polarization potentials is
\be
\Delta E_{HFS}^{2S} = 22.7690(5)~{\rm meV}.
\ee

We note that the effect of including the exponential finite-size
Coulomb potential as compared to the point case reduces the splitting
by 0.1269(5)~meV; introducing the point vacuum polarization potential
increases the splitting by 0.0747(5)~meV for the point-Coulomb, and
0.0742(5)~meV for the finite-Coulomb cases. Using the combined finite
vacuum polarization and finite-Coulomb potentials reduces the
splitting by 0.0012(5)~meV to give the value above.

Alternatively, one can follow Ref.~\cite{Borie:1982ax} in which case
we can calculate this splitting to be
\beqa
\Delta E_{HFS}^{2S} &=& \frac{\kappa g \mu}{\kappa^2-\frac{1}{4}}\, 
\left[\Lambda(\Lambda+1)-I(I+1)-j(j+1)\right]\notag \\
&&\times \frac{\alpha}{2M_p}\int r^{-2}g(r)f(r)dr.
\eeqa
In that case, we calculate 
\be
\Delta E_{HFS}^{2S} = 22.7640(5)~{\rm meV},
\ee
provided we correct for the reduced mass in the formula such that
the magnetic moment of the muon is not defined in terms of reduced
mass but rather defined in terms of the free space mass.

A comparison to perturbative calculations is given in
Table~\ref{tab:2shfs}, where we note the finding of a finite-size
dependent contribution in this splitting, which is neglected in the
summary of Pohl {\it et al.} (though finite-size effects sans a
parameterization are calculated in studies by
Borie~\cite{Borie:2004fv} and Pachucki~\cite{Pachucki:1996zza}) and
which differs from results obtained using the standard Zemach
treatment~\cite{Pachucki:1996zza}. We note that for $\bra
r_p^2\ket^{1/2} = 0.8768$~fm, the $2S_{1/2}$ hyperfine splitting is
calculated here to be 22.8496~meV (22.8547~meV for $\bra
r_p^2\ket^{1/2} = 0.84184$~fm) which indicates a $0.0087(5)$~meV
($0.0100(5)$~meV) correction to the perturbative calculation (once the
factor of 1/4 is taken into account), or $2.8$--$3.2\%$ of the 0.31~meV
quoted discrepancy.

%
\begin{center}
\begin{table*}[b]
\footnotesize
\caption{Contributions to the $2S_{1/2}$ hyperfine splitting
  calculated via Eq.~(\ref{eq:2shfs_weisskopf}) with comparison to
  values found in Martynenko~\cite{Martynenko:2004bt}. Subscripts are
  defined in Table~\ref{tab:lamb}. Values are all in meV. Errors in
  the Dirac calculations are taken to be $\pm 500$~neV as per
  Section~\ref{sec:qc}. The listed corrections are already included in
  our Dirac calculations and are listed by their descriptions in
  Ref.~\cite{Martynenko:2004bt}. All further corrections to both the
  perturbative calculation and our calculation are contained in
  `Remaining Corrections' which encompasses the muon AMM, amongst
  other corrections listed in Ref.~\cite{Martynenko:2004bt}. We note
  that the `Proton structure corrections of ${\cal O}(\alpha^5)$'
  pertains to the Zemach contribution (which we shall explore in an
  upcoming publication) and does not include considerations of
  finite-size in the wavefunction, and that the polynomial dependence
  on $\bra r_p^2\ket^n$ of this splitting is not discussed in the
  literature.  \label{tab:2shfs}\vspace{2mm}}
\begin{tabular}{l.c}
&& \\[-3mm]
\hline
\hline\\[-2mm]
Contribution & \multicolumn{1}{c}{\phantom{abc}Martynenko\phantom{abc}} & Present Work \\[1mm]
\hline\\[-2mm]

Dirac ($V = V_{\rm C}$) & & 22.8229 \\[1mm]
Dirac ($V = V_{\rm C} + V_{\rm VP}$) & & 22.8976 \\[1mm]
Dirac ($V = V_{\rm FC}$) & & 22.7774 - 0.1746 $\bra r_p^2\ket$ + 0.0709 $\bra r_p^2\ket^{3/2}$ \\[1mm]
Dirac ($V = V_{\rm FC} + V_{\rm VP}$) & & 22.8510 - 0.1701 $\bra r_p^2\ket$ + 0.0667 $\bra r_p^2\ket^{3/2}$ \\[1mm]
\hline
&& \\[-2mm]
Dirac ($V = V_{\rm FC} + V_{\rm FVP}$) & & 22.8521 - 0.1795 $\bra r_p^2\ket$ + 0.0739 $\bra r_p^2\ket^{3/2}$ \\[1mm]

Fermi Energy $E_F$ &
22.8054 &   \\[1mm]

Relativistic correction $\frac{17}{8}(Z\alpha)^2E_F$ & 
0.0026 &   \\[1mm]

VP corrections of orders $\alpha^5$, $\alpha^6$ & & \\[1mm]
in the second order of perturbation series &
0.0746 & \\[1mm]

Proton structure corrections of order $\alpha^5$ &
-0.1518 &   \\[1mm]

Proton structure corrections of order $\alpha^6$ &
-0.0017 &   \\[1mm]

\hline
&& \\[-2mm]
Subtotal: & 22.7291 & 22.8521 - 0.1795 $\bra
r_p^2\ket$ + 0.0739 $\bra r_p^2\ket^{3/2}$ \\[1mm]

\hline
&& \\[-2mm]

Remaining Corrections & \multicolumn{2}{c}{\phantom{abc}0.0857} \\[1mm]

\hline
&& \\[-2mm]
Total: & 22.8148 & 22.9378 - 0.1795 $\bra
r_p^2\ket$ + 0.0739 $\bra r_p^2\ket^{3/2}$ \\[1mm]

\hline
\hline

\end{tabular}
\end{table*}
\end{center}
%
%
\subsection{$2P_{1/2}$ Hyperfine Structure}\label{sec:2p1/2HFS}

The $2P_{1/2}$ hyperfine structure is of no consequence for the
transition which we are investigating here. Nonetheless, we calculate
the energy of the $2P_{1/2}^{F=0}$ and $2P_{1/2}^{F=1}$ levels as a
confirmation of our method, and to compare to the perturbative
results. Following Ref.~\cite{Martynenko:2006gz}, to ${\cal
  O}(\alpha^4)$ the $2P_{1/2}$ hyperfine structure splitting is given by
\be
\label{eq:2p12martynenko}
\Delta E_{HFS}^{{2P}_{1/2}} = 
E_F\left[\frac{1}{3}+\frac{a_\mu}{6}+\frac{m_\mu(1+2\kappa)}{12m_p(1+\kappa)}\right],
\ee
for which the Fermi energy is 
\be
\label{eq:ef}
E_F = \frac{\mu^3(1+\kappa)}{3m_\mu m_p}(Z\alpha)^4,
\ee
and where $\alpha_\mu$ is the muon anomalous magnetic moment. We
note another important, relevant typographical correction; in
Ref.~\cite{Martynenko:2006gz} the factors of 2 in the denominators of
the third terms of Eqs.~(27--28) should read 12. The calculations are
performed correctly however. For $\ell\neq 0$, the dipole term in the
Hamiltonian (Eq.~(\ref{eq:hamiltonian})) is non-zero, while the
contact term vanishes. The energy for the dipole term is thus given by
\be
E_{\rm HFS}^{{2P}_{1/2}} = 2\beta\gamma\hbar\frac{\ell(\ell+1)}{j(j+1)}\bigg\bra\frac{1}{r^3}\bigg\ket
  \bra Fm_F| \mathbf{I}\cdot\mathbf{J} | Fm_F \ket,
\ee
where the non-zero terms in the dot-product are given by
\be \label{eq:idotj}
\bra Fm_F| \mathbf{I}\cdot\mathbf{J} | Fm_F \ket = \half [F(F+1)-I(I+1)-j(j+1)].
\ee
For Schr\"odinger wave functions, the vacuum expectation value
of $r^{-3}$ is analytic, in that
\be
\bigg\bra\frac{1}{r^3}\bigg\ket = \left(a_0^3 n^3 \ell(\ell+1)(\ell+\half)\right)^{-1}.
\ee
Inserting the appropriate values of $F, n, \ell, I, j$ for each of
the $F=0$ and $F=1$ states, one obtains the energy of the $2P_{1/2}$
hyperfine structure to be
\be
\label{eq:2p12weissbluth}
\Delta E_{\rm HFS}^{{2P}_{1/2}} =
\frac{2}{9}\beta\gamma\hbar/a_0^3\, .
\ee
Eq.~(\ref{eq:2p12weissbluth}) corresponds to the leading term of
Eq.~(\ref{eq:2p12martynenko}), to which the anomalous magnetic moments
provide additional corrections. Using the converged Dirac
wave functions with exponential finite-Coulomb and finite vacuum
polarization potentials (rather than Schr\"odinger wave functions) we
calculate the expectation value of $r^{-3}$ and find
\be
\Delta E_{\rm HFS}^{{2P}_{1/2}} = 7.6204(5)~{\rm meV}.
\ee
The results of this calculation are summarised in
Table~\ref{tab:2p12hfs} where we note that the addition of the (point)
vacuum polarization potential to the point-Coulomb potential increases
the splitting by 0.0017(5)~meV, and the introduction of the
finite-Coulomb potential increases this further by 0.0045(5)~meV to
arrive at the value above. The effect of finite-vacuum polarization is
essentially zero here.

%
\begin{center}
\begin{table}[bt]
\footnotesize
\caption{Contributions to the $2P_{1/2}$ hyperfine splitting with
  comparison to values found in
  Martynenko~\cite{Martynenko:2006gz}. Subscripts are defined in
  Table~\ref{tab:lamb}. Values are all in meV. Errors in the Dirac
  calculations are taken to be $\pm 500$~neV as per
  Section~\ref{sec:qc}. The listed corrections are already included in
  our Dirac calculations. The `Leading contribution' is extracted from
  the ${\cal O}(Z\alpha)^4$ line in Table II of
  Ref.~\cite{Martynenko:2006gz} and includes only the $E_F/3$ term of
  Eq.~(\ref{eq:2p12martynenko}). The relativistic correction is listed as
  ${\cal O}(Z\alpha)^6$ in that reference. All further corrections to
  both the perturbative calculation and our calculation are contained
  in `Remaining Corrections' which encompasses the muon AMM, proton
  MM, and all other lines of Table II in Ref.~\cite{Martynenko:2006gz}
  not already listed. We note the minor error in the last digit of
  the summary contribution of Ref.~\cite{Martynenko:2006gz} for this
  state, likely arising from round-off
  error. \label{tab:2p12hfs}\vspace{2mm}}
\begin{tabular}{l..}
&& \\[-3mm]
\hline
\hline\\[-2mm]
Contribution & \multicolumn{1}{c}{\phantom{abc}Martynenko\phantom{abc}} & \multicolumn{1}{c}{\phantom{abc}Present Work\phantom{abc}} \\[1mm]
\hline\\[-2mm]

Dirac ($V = V_{\rm C}$) & & 7.6141 \\[1mm]
Dirac ($V = V_{\rm C} + V_{\rm VP}$) & & 7.6159 \\[1mm]
Dirac ($V = V_{\rm FC} + V_{\rm VP}$) & & 7.6204 \\[1mm]
\hline
&& \\[-2mm]
Dirac ($V = V_{\rm FC} + V_{\rm FVP}$) & & 7.6204 \\[1mm]

Leading contribution &
7.6018 & \\[1mm]

${\cal O}(Z\alpha)^6$ contribution & 0.0011 & \\[1mm]

\hline
&& \\[-2mm]
Subtotal &
7.6029 & 7.6204 \\[1mm]

\hline
&& \\[-2mm]
Remaining Corrections & \multicolumn{2}{c}{\phantom{abc}0.3615} \\[1mm]

\hline
&& \\[-2mm]
Total: & 7.9644 & 7.9819 \\[1mm]

\hline
\hline

\end{tabular}
\end{table}
\end{center}
%

%
\subsection{$2P_{3/2}$ Hyperfine Structure}\label{sec:2p3/2HFS}

Following the same method as in the previous subsection, we can
calculate the energy levels for the $2P_{3/2}^{F=1}$ and
$2P_{3/2}^{F=2}$ levels. The $2P_{3/2}$ hyperfine structure, as
derived in Ref.~\cite{Martynenko:2006gz} is given by
\be
\Delta E_{HFS}^{{2P}_{3/2}} = 
E_F \left[\frac{2}{15}-\frac{a_\mu}{30}+\frac{m_\mu(1+2\kappa)}{12m_p(1+\kappa)}\right],
\ee
where $E_F$ is given in Eq.~(\ref{eq:ef}) (we note again the
typographical correction detailed in
Section~\ref{sec:2p1/2HFS}). Alternatively, inserting the relevant
values for this state into Eq.~(\ref{eq:idotj}) and using the
converged Dirac wave functions we find
\be
\Delta E_{\rm HFS}^{{2P}_{3/2}} = 3.0415(5)~{\rm meV}
\ee
when the potential consists of the exponential finite-Coulomb and
point vacuum polarization potentials. For this state, the addition of
the (point) vacuum polarization potential to the point-Coulomb
potential increases the splitting by 0.0007(5)~meV to the value listed
in Table~\ref{tab:2p32hfs}, and the introduction of the finite-Coulomb
potential was found to have no effect within the limits of our
calculation, so too was the introduction of the finite vacuum
polarization potential.

%
\begin{center}
\begin{table}
\footnotesize
\caption{Contributions to the $2P_{3/2}$ hyperfine splitting with
  comparison to values found in
  Martynenko~\cite{Martynenko:2006gz}. Subscripts are defined in
  Table~\ref{tab:lamb}. Values are all in meV. Errors in the Dirac
  calculations are taken to be $\pm 500$~neV as per
  Section~\ref{sec:qc}. The details of the listed corrections are the
  same as those of
  Table~\ref{tab:2p12hfs}. \label{tab:2p32hfs}\vspace{2mm}}
\begin{tabular}{l..}
&& \\[-3mm]
\hline
\hline\\[-2mm]
Contribution & \multicolumn{1}{c}{\phantom{abc}Martynenko\phantom{abc}} & \multicolumn{1}{c}{\phantom{abc}Present Work\phantom{abc}} \\[1mm]
\hline\\[-2mm]

Dirac ($V = V_{\rm C}$) & & 3.0408 \\[1mm]
Dirac ($V = V_{\rm C} + V_{\rm VP}$) & & 3.0415 \\[1mm]
Dirac ($V = V_{\rm FC} + V_{\rm VP}$) & & 3.0415 \\[1mm]
\hline
&& \\[-2mm]
Dirac ($V = V_{\rm FC} + V_{\rm FVP}$) & & 3.0415 \\[1mm]

Leading contribution &
3.0407 & \\[1mm]

Relativistic correction &
0.0001 & \\[1mm]

\hline
&& \\[-2mm]
Subtotal &
3.0408 & 3.0415 \\[1mm]

\hline
&& \\[-2mm]
Remaining Corrections & \multicolumn{2}{c}{\phantom{abc}0.3518} \\[1mm]

\hline
&& \\[-2mm]
Total: & 3.3926 & 3.3933 \\[1mm]

\hline
\hline

\end{tabular}
\end{table}
\end{center}
%

%
\section{Summary}\label{sec:summary}

We summarise the findings of these non-perturbative Dirac calculations
and compare to the previous literature values of perturbative
calculations in Table~\ref{tab:summary}. We add to this a combined
expression for the cubic which when set equal to the experimental
value of the measured transition is solved to predict the proton rms
charge-radius, as is done in Ref.~\cite{Pohl:2010zz}.

We do not include the result of hyperfine splitting calculations for
the $2P_{1/2}$ eigenstate as this is of no relevance to the measured
transition. We also note the omission here of the energy shift
attributed to a mixing between the $2P_{1/2}$ and $2P_{3/2}$ $F=1$
states, as discussed in Ref.~\cite{Pachucki:1996zza} for comparison to
Ref.~\cite{Pohl:2010zz} where it is also absent.

\begin{center}
\begin{table*}[!h]
\footnotesize
\caption{Sum of perturbative and non-perturbative theoretical
  contributions to the measured experimental transition energy shown
  in Fig.~\ref{fig:spectrum}. Subscripts are defined in
  Table~\ref{tab:lamb}. Values are all in meV. The individual
  perturbative contributions (listed under `Various') are taken from
  Tables~\ref{tab:lamb}--\ref{tab:2p32hfs}. In each case,
  the value given for the Dirac calculation is calculated using the
  combination of finite-Coulomb and finite vacuum polarization
  potentials ($V=V_{\rm FC}+V_{\rm FVP}$). The fractional factors for
  the hyperfine splittings are inserted for relevance to the measured
  transition, and are calculated via angular-momentum splitting
  rules.\vspace{2mm}\label{tab:summary}}
\begin{tabular}{l..}
&& \\[-3mm]
\hline
\hline\\[-2mm]
Contribution & \multicolumn{1}{c}{Various} &
\multicolumn{1}{c}{Present Work}  \\[1mm]
\hline\\[-2mm]

$2S_{1/2}$-$2P_{1/2}$ Lamb shift (constant) & 206.0573  & 206.0604 \\[1mm]

$2S_{1/2}$-$2P_{1/2}$ Lamb shift (finite-size)\phantom{abcdefg} &
\multicolumn{1}{c}{$\fixminus 5.2262 \bra r_p^2\ket + 0.0347 \bra r_p^2\ket^{3/2}$}&
\multicolumn{1}{c}{$\fixminus 5.2794 \bra r_p^2\ket + 0.0546 \bra r_p^2\ket^{3/2}$} \\[1mm]

$2P$ Fine Structure & 8.3521 & 8.3521 \\[1mm]

$\frac{1}{4} \times 2S_{1/2}$ Hyperfine (constant) & -5.7037 & -5.7345 \\[1mm]
$\frac{1}{4} \times 2S_{1/2}$ Hyperfine (finite-size) & 0.0000 & \multicolumn{1}{c}{0.0449 $\bra r_p^2\ket$ $-$ 0.0185 $\bra r_p^2\ket^{3/2}$} \\[1mm]

$\frac{3}{8} \times 2P_{3/2}$ Hyperfine & 1.2722 & 1.2725  \\[1mm]

\hline
\hline
&& \\[-1mm]
{\bf \ \ \  (Various) Total (Perturbative)} & 
\multicolumn{2}{c}{\bf 209.9779 
- 5.2262 $\mathbf{\bra r_p^2\ket}$
+ 0.0347 $\mathbf{\bra r_p^2\ket^{3/2}}$} \\[2mm]
&& \\[-1mm]
{\bf \ \ \  (Present Work) Total (Dirac)} & 
\multicolumn{2}{c}{\bf 209.9505
- 5.2345 $\mathbf{\bra r_p^2\ket}$ 
+ 0.0361 $\mathbf{\bra r_p^2\ket^{3/2}}$}  \\[2mm]

\hline
\hline

\end{tabular}
\end{table*}
\end{center}

We find that the perturbative calculations are largely reproduced
using our methods when considering the appropriate potentials for
comparison. We further find that in several cases the use of the
finite-vacuum polarization potential produces effects which are not
accounted for in previous studies. The largest of these is the
finite-size contribution to the $2S_{1/2}$ hyperfine splitting which
has been neglected in the literature up to this point.

Overall, the non-perturbative calculations do not elucidate any
missing contributions of a magnitude large enough to resolve the
proton radius problem outlined in Pohl {\it et
  al.}~\cite{Pohl:2010zz}.


%
\section{Conclusions}\label{sec:conclusions}

After careful consideration of the various contributions to the
measured transition energy of Pohl {\it et al.}~\cite{Pohl:2010zz},
calculated consistently using the Dirac equation with appropriate
potentials, and following the addition of the required corrections to
these calculations (taking further care to avoid overcounting issues),
we find no single term which leads to a discrepancy with the
perturbative results of sufficient magnitude to account for the
discrepancy reported in Ref.~\cite{Pohl:2010zz}. These calculations
nonetheless provide a useful insight into the reliability of the
perturbative calculations, and allow a simpler approach to future
investigations.

While it remains possible in principle that one or more of the
higher-order corrections to the terms calculated in this work might be
of sufficient magnitude to affect the analysis of
Ref.~\cite{Pohl:2010zz}, the precision with which the Dirac and
perturbative calculations agree for the terms which we have calculated
here strongly suggests that this will not be the case.

In addition to the calculations presented here, we further note that
our calculations of a hitherto overlooked contribution arising from
off-mass-shell effects for the proton (which are negligible for
electronic hydrogen) provide a natural solution to the proton radius
problem~\cite{Miller:2011yw}, and as such the combination of these two
sets of calculations may be seen as a complete description of the
measured transition in muonic hydrogen with no discrepancy in the rms
charge radius of the proton. Because of the uncertain magnitude of the
off-mass-shell effects, it is incorrect to complete the analysis of
this transition to predict a proton rms charge radius\emdash we await
the results of current and future experiments which will be ascertain
the strength of this contribution, after which a complete analysis
will be possible.

Nonetheless, we note that our calculations predict that the transition
energy for the $2P_{3/2}^{F=2}$ to $2S_{1/2}^{F=1}$ transition in
muonic hydrogen is larger in magnitude than that which is predicted
by the perturbative calculations, and that analysis of this data under
the assumption that no further terms are required leads to the
following values for the proton rms charge radius when fit to the
experimental data;
\bea
\nonumber
&{\bf Pohl\ {\it et\ al.}}:\ &\sqrt{\bra r_p^2\ket} = 0.84183(67)~{\rm fm}, \\[2mm]
\nonumber
&{\bf Present\ Work}:\ &\sqrt{\bra r_p^2\ket} = 0.83811(67)~{\rm fm}.
\eea
The value listed as Present Work is taken as the solution to the cubic
equation
\be \label{eq:solve}
209.9505 - 5.2345 \bra r_p^2\ket + 0.0361 \bra r_p^2\ket^{3/2} = 206.2949,
\ee
where the right-hand-side corresponds to the quoted value of the
measured transition in Ref.~\cite{Pohl:2010zz}; the left-hand-side is
taken from the relevant conclusion line of Table~\ref{tab:summary};
and for which the errors in this calculation are dominated by the
experimental error. The extracted $\sqrt{\bra r_p^2\ket}$ value listed
as Pohl {\it et al.} is taken from Ref.~\cite{Pohl:2010zz} (calculated
in the same fashion) and differs from the central value quoted
in~\cite{Pohl:2010zz}; 0.84184(67), though the difference is well
within the quoted errors.

We note the degree to which the cubic expression Eq.~(\ref{eq:solve})
agrees with that of Ref.~\cite{Pohl:2010zz}, despite the latter not
involving a calculation of a finite-size contribution to the $2S$
hyperfine splitting. Some research in progress by the authors will
elucidate some further overlooked contributions that will likely alter
the agreement between these two expressions, and we look forward to
future measurements with which we may compare our findings.

%
\begin{acknowledgments}

This research was supported in part by the United States Department of
Energy (under which Jefferson Science Associates, LLC, operates
Jefferson Lab) via contract DE-AC05-06OR23177 (JDC, in part); grant
FG02-97ER41014 (GAM); and grant DE-FG02-04ER41318 (JR), and by the
Australian Research Council (through grant FL0992247) and the
University of Adelaide (JDC, AWT). GAM and JR gratefully acknowledge
the support and hospitality of the University of Adelaide while the
project was undertaken. The authors thank R. Pohl, E. Borie, and
F. Kottmann for their comments and corrections of numerical and
typographical errors.

\end{acknowledgments}
%
%
\bibliography{muHrefs}
\end{document}